%% file: gwpasan.tex
\documentclass[sigconf,natbib=false,nonacm,authorversion]{acmart}

\usepackage[english]{babel}
\usepackage[T1]{fontenc}
\usepackage[utf8]{inputenc}

\usepackage{microtype}
\usepackage{array}
\usepackage{calc}
\usepackage{hyperref}
\usepackage[all]{hypcap} 
\usepackage{amsmath}
\usepackage[page]{appendix}
\usepackage{graphicx}
\usepackage{float}
\usepackage{enumerate}
\usepackage{enumitem}
\setlist{noitemsep, leftmargin=*, topsep=0.5ex, partopsep=0.5ex}

\usepackage{listings}
\lstdefinelanguage
   [x64]{Assembler}
   [x86masm]{Assembler}
   {morekeywords={
      fisttpll,
      fldln2,
      fprem1,
      fld1,
      fcos,
      f2xm1,
      rax, rdi
   }}

\usepackage{tikz}
\usetikzlibrary{patterns}
\usetikzlibrary{positioning}

\usepackage[bottom]{footmisc}

\usepackage[backend=biber,
  style=acmnumeric,
  datamodel=acmdatamodel,
  sorting=none, 
  maxbibnames=99,
]{biblatex}
\usepackage{csquotes} 
\bibliography{ref}

\newcommand{\runinsec}[1]{\vspace{0.2em}\noindent\textbf{#1} }
\newcommand{\inlcode}[1]{\textsf{#1}}

\title{GWP-ASan: Sampling-Based Detection of Memory-Safety Bugs in Production}

\author{Kostya Serebryany}
\affiliation{\institution{Google}\city{}\country{}}
\email{kcc@google.com}

\author{Chris Kennelly}
\affiliation{\institution{Google}\city{}\country{}}
\email{ckennelly@google.com}

\author{Mitch Phillips}
\affiliation{\institution{Google}\city{}\country{}}
\email{mitchp@google.com}

\author{Matt Denton}
\affiliation{\institution{Google}\city{}\country{}}
\email{mpdenton@google.com}

\author{Marco Elver}
\affiliation{\institution{Google}\city{}\country{}}
\email{elver@google.com}

\author{Alexander Potapenko}
\affiliation{\institution{Google}\city{}\country{}}
\email{glider@google.com}

\author{Matt Morehouse}
\authornote{At Google when working on this project.}
\affiliation{%
  \institution{Independent Researcher}
  \city{}
  \country{}
}
\email{mattmorehouse@gmail.com}

\author{Vlad Tsyrklevich}
\authornotemark[1]
\affiliation{%
  \institution{Independent Researcher}
  \city{}
  \country{}
}
\email{vlad@tsyrklevi.ch}

\author{Christian Holler}
\affiliation{\institution{Mozilla Corporation}\city{}\country{}}
\email{choller@mozilla.com}

\author{Julian Lettner}
\affiliation{\institution{Apple}\city{}\country{}}
\email{julian.lettner@apple.com}

\author{David Kilzer}
\affiliation{\institution{Apple}\city{}\country{}}
\email{ddkilzer@apple.com}

\author{Lander Brandt}
\affiliation{\institution{Meta}\city{}\country{}}
\email{landerb@meta.com}

\date{}

\begin{document}
\input{tex/abstract.tex}

\ccsdesc[500]{Software and its engineering~Allocation / deallocation
strategies}
\ccsdesc[500]{Security and privacy~Software security engineering}
\ccsdesc[500]{Theory of computation~Program analysis}

\keywords{Memory Safety, Dynamic Program Analysis, Programming Languages,
Software Engineering}

\maketitle


\input{tex/intro.tex}
\input{tex/background.tex}
\input{tex/gwpasan.tex}
\input{tex/eval.tex}
\input{tex/related.tex}
\input{tex/future.tex}
\input{tex/conclusion.tex}

\printbibliography



\end{document}

%% file: tex/abstract.tex
\begin{abstract}

  Despite the recent advances in pre-production bug detection,
  heap-use-after-free and heap-buffer-overflow bugs remain the primary problem
  for security, reliability, and developer productivity for applications
  written in C or C++, across all major software ecosystems. Memory-safe
  languages solve this problem when they are used, but the existing code bases
  consisting of billions of lines of C and C++ continue to grow, and we need
  additional bug detection mechanisms.

  This paper describes a family of tools that detect these two classes of
  memory-safety bugs, while running in production, at near-zero overhead. These
  tools combine page-granular guarded allocation and low-rate sampling. In
  other words, we added an ``if'' statement to a 36-year-old idea and made it
  work at scale.

  We describe the basic algorithm, several of its variants and implementations,
  and the results of multi-year deployments across mobile, desktop, and server
  applications.

\end{abstract}

%% file: tex/intro.tex
\section{Introduction}

Memory-safety bugs have been well-known since at least
1972~\cite{Anderson1972}.  Exploitation of memory unsafety made news in
1988~\cite{Orman2003} and ever since.  Numerous dynamic-analysis-based
detection mechanisms for pre-production use have been implemented: Valgrind
Memcheck~\cite{NethercoteS2007}, AddressSanitizer
(ASan)~\cite{SerebryanyBPV2012}, and HardwareAddressSanitizer
(HWASan)~\cite{SerebryanySSTV2018} being the most popular. At least two
hardware detection mechanisms have been introduced; namely SPARC
ADI~\cite{AingaranJKLLMPR2015}, which has disappeared alongside its hardware
platform, and Arm MTE~\cite{Serebryany2019}, which is not yet widely available.
The devastating impact of memory unsafety was one of the reasons for the
creation of newer and safe(r) languages: Java, C\#, Go, Swift, and Rust.
Advancements in automated pre-production testing and fuzzing (``shift left''
testing~\cite{Smith2001}) lead to detection and elimination of millions of
memory-safety bugs.  Yet they remain the single major source of security
vulnerabilities, and continue to negatively impact reliability and developer
productivity~\cite{SzekeresPWS2013, Gaynor2020}.

In the meantime, perhaps the oldest known detection mechanism remained
underutilized. The Electric Fence Malloc Debugger~\cite{EFence}, introduced in
1987, detects heap-use-after-free and heap-buffer-overflow bugs. It replaces
the standard \inlcode{malloc()} and \inlcode{free()} functions. The
\inlcode{malloc()} function rounds up the allocation size to the virtual memory
system's page size, allocates the required size plus two extra pages directly
via the \inlcode{mmap()} system call, and uses the \inlcode{mprotect()} system
call to make the first and the last page inaccessible. These pages are called
the ``redzone'' or ``guard pages''. The address of the first unprotected page
is returned to the user. The original implementation uses only one guard page
at a time, but other variants use two.  The \inlcode{free()} function calls
\inlcode{mprotect()} on the entire memory region and prevents this virtual
address range from being reused soon. Any memory access to the guard page or
the deallocated address range causes a segmentation fault, at no additional
effort.

Elegant and easy to use, this mechanism suffers from high execution costs.
Rounding up the allocation size to the page granularity may cause \(100\times\)
RAM overhead, while one system call per allocation and deallocation causes
slowdowns of comparable magnitude. Electric Fence~\cite{EFence} and its many
clones~\cite{PageHeap} remain unusable outside of small tests.

The tool named \emph{GWP-ASan} and its variants, first introduced in 2018, adds
an ``if'' statement to the Electric Fence algorithm and makes it a successful
sampling-based bug detector for production use, with amortized near-zero
overhead. Today several implementations of GWP-ASan run in production in
mobile, desktop, and server ecosystems. They have detected thousands of
memory-safety bugs in production that evaded all other kinds of detection.
GWP-ASan is a feature inside of \inlcode{malloc()} implementations and does
not require any modifications to program binaries.

GWP-ASan does not replace tools like ASan or HWASan for regular pre-production
testing due to its extremely low probability of detecting a bug, per instance.
However, the low probability of per-instance detection is offset by large-scale
production deployment, with the aggregate detection probability resulting in a
large number of detections. The detailed error messages usually enable
developers to fix the bugs without the reproducers being available.

We do not know how many memory-safety bugs remain undetected. GWP-ASan finds
the bugs that occur in production \emph{frequently} and misses most others.
GWP-ASan is \emph{not a security mitigation} tool due to its low detection
probability.

The name ``GWP-ASan'' is derived from Google-Wide Profiling
(GWP)~\cite{RenTMSRH2010}---a tool that collects profiling data by, amongst
other things, sampling \inlcode{malloc()}---and AddressSanitizer
(ASan)~\cite{SerebryanyBPV2012}---a tool that detects use-after-frees and
heap-buffer-overflows---even though GWP-ASan is neither GWP nor ASan. Three
independent implementations of this tool use this name, and other
implementations are named differently.

\runinsec{Paper organization.} The following section, \S\ref{sec:background},
will introduce background on the types of bugs that GWP-ASan can detect;
\S\ref{sec:algo} describes the high-level GWP-ASan algorithm design;
\S\ref{sec:impls} describes several implementations of that algorithm for a
variety of platforms; \S\ref{sec:results} discusses real-world results from
deployment of the implementations; \S\ref{sec:related} and \S\ref{sec:future}
discuss related work and opportunities for future work.

%% file: tex/background.tex
\section{Background}
\label{sec:background}

Before diving deeper, we will take a brief look at the types of bugs that
GWP-ASan will be able to detect, and trade-offs in dynamic program analysis.

\runinsec{Heap memory-safety bugs.} Fundamentally, memory-safety is a property
of a programming language. Different languages choose different strategies for
memory safety~\cite{SzekeresPWS2013}. Unsafe languages, specifically the C and
C++ programming languages, define some well-typed programs to have
\emph{undefined behavior}, the source of which are considered bugs in the
program. Heap buffer overflows (viz. out-of-bounds accesses) and use-after-free
accesses (viz. dangling-pointer accesses) are two such bugs.

A \emph{heap buffer overflow} happens when an object of a certain size is
allocated on the heap, and then a pointer to this object is used to access
memory outside of the object's bounds. Typically, the object is an array of $n$
elements, and the code accesses the $i$-th element where $i < 0$ or $i \geq n$.
Listing~\ref{lst:oob} shows examples of buffer overflows in C code. Modern
compilers can provide warnings for simple buffer overflows, including the ones
shown in the example. Unfortunately, the non-obvious cases that also evade
human review are far more common.

\vspace{-1em}
\begin{lstlisting}[language=C, frame=single, xleftmargin=0.5em,
  xrightmargin=0.5em, caption=Buffer overflow examples., label=lst:oob]
// heap allocation
int *array = malloc(n * sizeof(int));
// buffer overflow
array[n] = 42;
// buffer overflow (underflow)
array[-1] = 42;
// buffer overflow, assuming n <= 100500
array[100500] = 42;
\end{lstlisting}

A \emph{heap use-after-free} happens when an object is allocated on the heap,
and later deallocated, while a pointer to the object is preserved elsewhere and
is used to access the deallocated memory after the deallocation.
Listing~\ref{lst:uaf} shows an example of a simple use-after free. Again,
modern compilers will provide warnings for simple cases as shown, but the
non-obvious cases are much more common and difficult to find (esp. cases
involving racy use-after-frees).

\begin{lstlisting}[language=C, frame=single, xleftmargin=0.5em,
xrightmargin=0.5em, caption=Use-after-free example., label=lst:uaf]
// heap allocation
int *val = malloc(sizeof(int));
// heap deallocation
free(val);
// heap use-after-free
*val = 0;
\end{lstlisting}

In both cases, the buggy memory access touches memory not belonging to the
respective objects. In the C and C++ standards, this is considered undefined
behavior. In practice, however, this may result in a crash, a silent data
corruption, or an exploitable security vulnerability~\cite{SzekeresPWS2013}.

\runinsec{Dynamic analysis.} Dynamic analysis tools perform program analysis at
runtime, observing state changes by instrumenting relevant instructions and
functions as the program is running on real inputs. More complex dynamic
analyses also maintain \emph{shadow state}, which uses additional memory or
metadata to maintain additional information about the program's state that is
not efficiently available otherwise (e.g. if memory is allocated or freed).

Consequently, typical dynamic analysis can only observe the program
transitioning into \emph{erroneous states}.  Consider the taxonomy defined by
Randell~\cite{Randell2003}, with a \emph{fault}---or simply ``bug''---being a
flaw in the program's logic, causing \emph{errors} which are bad states that
can ultimately lead to program \emph{failure}, i.e. the program crashes, cannot
deliver the requested service or worse. One of the biggest challenges in
dynamic analysis is giving developers the information to localize faults,
despite only being able to detect the resulting error. Indeed, producing
helpful reports that allow developers to debug and fix a bug requires
maintaining additional information (e.g. stack traces of where memory was
allocated or freed), which can be very costly.

Analyzing memory accesses, as required for memory-safety analysis, adds
additional overheads, depending on the precision of the analysis. Tools such as
AddressSanitizer choose to analyze every memory access, along with maintaining
every allocation's and deallocation's metadata (valid, invalid, size, stack
trace).  Unsurprisingly, this results in significant runtime and memory
overheads, which make such approaches unsuitable for production environments.

Hardware-based solutions make use of new CPU features, such as Arm
MTE~\cite{Serebryany2019}, to offload storing some or all of the additional
metadata and checking to dedicated hardware. Unfortunately, some of the newer
hardware-based solutions are not yet widely available. GWP-ASan makes use of a
hardware feature available in all modern CPUs' Memory Management Units
(MMUs)~\cite{CAR}: paged virtual memory and the ability to set memory pages
inaccessible.

%% file: tex/gwpasan.tex
\section{GWP-ASan Algorithm}
\label{sec:algo}

This section describes the basic GWP-ASan algorithm design. Given all
implementations discussed in \S\ref{sec:impls} are either implemented in the C
or C++ languages, the code snippets in this section show simplified C language
snippets. For the description of our implementation, we assume a
POSIX-compatible operating system.

We define the original pseudo-implementation of the heap allocator in
Listing~\ref{lst:original}; the details of the original algorithms for heap
allocation, in \inlcode{Allocate()}, and deallocation, in
\inlcode{Deallocate()}, can be treated as a black box for the simple algorithm
described here.

\begin{lstlisting}[language=C, frame=single, xleftmargin=3em,
  xrightmargin=3em, caption=Unmodified heap allocation., label=lst:original]
  void *malloc(size_t size) {
    return Allocate(size);
  }

  void free(void *ptr) {
    Deallocate(ptr);
  }
\end{lstlisting}

In order to implement sampling heap error detection, the change as shown in
Listing~\ref{lst:sampling} should be made.

\begin{lstlisting}[language=C, frame=single, xleftmargin=3em,
  xrightmargin=3em, caption=Heap allocation with sampling error detection.,
  label=lst:sampling, escapechar=@]
  void *malloc(size_t size) {
@$\oplus$@   if (WantToSample(size))
@$\oplus$@     return GuardAlloc(size);
    return Allocate(size);
  }

  void free(void *ptr) {
@$\oplus$@   if (IsGuarded(ptr)) {
@$\oplus$@     GuardDealloc(ptr);
@$\oplus$@     return;
@$\oplus$@   }
    Deallocate(ptr);
  }
\end{lstlisting}

\inlcode{WantToSample()} returns true infrequently. \inlcode{GuardAlloc()}
allocates, and \inlcode{GuardDealloc()} deallocates memory similar to Electric
Fence.  \inlcode{IsGuarded()} returns true if its argument is a pointer
previously returned by \inlcode{GuardAlloc()}.  Details vary between
implementations and some of them are described below.

\subsection{Simple Version}

In this section we describe a very simple, yet fully capable, implementation of
GWP-ASan.

\runinsec{Initialization.} At process start-up the GWP-ASan pool is allocated
using \inlcode{mmap()} as a fixed-size region of virtual memory consisting of
\(N\) page-sized allocation slots and \(N+1\) page-sized guards, such that
guards on both ends surround every slot. Figure~\ref{fig:initial} illustrates
the initial memory state of the GWP-ASan pool. Initially all of this memory is
marked as inaccessible (with \inlcode{PROT\_NONE}). The red zones remain
inaccessible throughout the program execution.

\begin{lstlisting}[language=C++, frame=single, xleftmargin=0.5em,
  xrightmargin=0.5em, caption=Basic implementation of \inlcode{WantToSample()}., label=lst:want_to_sample]
// Returns a random number in the range
// [1 ... sample_rate * 2].
int RandSkip();

// Initialized to a non-zero random number
// at thread start up.
static thread_local int skip = RandSkip();

bool WantToSample() {
  if (--skip> 0) return false;
  skip = RandSkip();
  return true;
}
\end{lstlisting}

\runinsec{Sampling allocations.} A possible implementation of
\inlcode{WantToSample()} is shown in Listing~\ref{lst:want_to_sample}. A
thread-local allocation \emph{skip} counter is used to decide which allocations
to skip sampling (the common case), and which to sample. The counter is
initialized to a random number in the range \([1 \ldots \inlcode{sample\_rate}
* 2]\), so that the median value is \inlcode{sample\_rate}. The skip counter is
then decremented by one per unsampled allocation. When the number reaches zero,
we try to service the \inlcode{malloc()} through GWP-ASan, and regenerate a new
random number.  Therefore, \inlcode{WantToSample()} in the common case is just
a single thread-local decrement and conditional branch.

\runinsec{Guarded allocation.} \inlcode{GuardAlloc()} checks if there are any
allocation slots available, chooses one, makes it accessible (with
\inlcode{PROT\_READ|PROT\_WRITE}), and returns to the caller.
Figure~\ref{fig:alloc} illustrates the pool's memory state after an allocation.

This way, buffer underflows (accesses below the allocation address by up to a
page) will always be detected, but overflows (accesses above the end of the
allocated region) will be detected only if the access crosses the page
boundary. To find overflows, the allocated region needs to be aligned right by
the page boundary. The implementation may randomly choose to align left or
right.

\runinsec{Deallocation of sampled allocations.} \inlcode{GuardDealloc()} marks
the allocation slot as inaccessible (with \inlcode{PROT\_NONE}). For
heap-use-after-free bugs to be detected with high probability, this
allocation slot needs to remain unallocated for some amount of time.
\inlcode{IsGuarded()} can be implemented as a range check.

\runinsec{Limitations.} The obvious limitations of this simple implementation
are:
\begin{enumerate}

  \item Only allocations of up to 1 page can be guarded.

  \item Only up to $N$ allocations can be guarded at the same time. For large
    enough $N$ this stops being a significant limitation because we want to
    avoid more than a certain number of guarded allocations anyway, to avoid
    high overheads.

\end{enumerate}

\DeclareRobustCommand{\redzone}{%
  \raisebox{-1pt}{\tikz \path [draw=gray,pattern={crosshatch}, pattern color=red] (0,0) rectangle (0.4,0.2);}
}
\DeclareRobustCommand{\unalloc}{%
  \raisebox{-1pt}{\tikz \path [draw=gray,fill=black] (0,0) rectangle (0.4,0.2);}
}
\DeclareRobustCommand{\alloc}{%
  \raisebox{-1pt}{\tikz \path [draw=gray,pattern={dots}, pattern color=green] (0,0) rectangle (0.4,0.2);}
}
\begin{figure}
  \begin{tikzpicture}
    \path [draw=gray,pattern={crosshatch}, pattern color=red] (0,0) rectangle (1,0.6);
    \path [draw=gray,fill=black] (1,0) rectangle (2,0.6);
    \path [draw=gray,pattern={crosshatch}, pattern color=red] (2,0) rectangle (3,0.6);
    \path [draw=gray,fill=black] (3,0) rectangle (4,0.6);
    \path [draw=gray,pattern={crosshatch}, pattern color=red] (4,0) rectangle (5,0.6);
    \path [draw=gray,fill=black] (5,0) rectangle (6,0.6);
    \path [draw=gray,pattern={crosshatch}, pattern color=red] (6,0) rectangle (7,0.6);
  \end{tikzpicture}
  \caption{GWP-ASan pool initial memory state. Guard pages, viz. ``red zones'',
  are shown as \redzone (red hatch pattern), and remain always inaccessible.
  Allocation slots are shown as \unalloc (black filled) when inaccessible.}
  \label{fig:initial}
\end{figure}
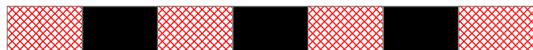
\begin{figure}
  \begin{tikzpicture}
    \path [draw=gray,pattern={crosshatch}, pattern color=red] (0,0) rectangle (1,0.6);
    \path [draw=gray,pattern={dots}, pattern color=green] (1,0) rectangle (2,0.6);
    \path [draw=gray,pattern={crosshatch}, pattern color=red] (2,0) rectangle (3,0.6);
    \path [draw=gray,fill=black] (3,0) rectangle (4,0.6);
    \path [draw=gray,pattern={crosshatch}, pattern color=red] (4,0) rectangle (5,0.6);
    \path [draw=gray,fill=black] (5,0) rectangle (6,0.6);
    \path [draw=gray,pattern={crosshatch}, pattern color=red] (6,0) rectangle (7,0.6);
  \end{tikzpicture}
  \caption{GWP-ASan pool memory state after an allocation. The allocation slot
  shown as \alloc (green dots) is allocated and accessible.}
  \label{fig:alloc}
\end{figure}
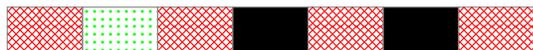

\subsection{Generating Descriptive Error Messages}

Generating descriptive error messages is crucial for GWP-ASan to be maximally
useful, given it is designed to run in production, where reproduction of any
given error is rather challenging. Without an easy way to reproduce an error,
developers require detailed information about detected errors to debug.

When a memory access hits a protected page, the operating-system kernel raises
a signal (\inlcode{SIGSEGV} on POSIX systems). If this page is a GWP-ASan guard
or a deallocated allocation slot, the signal handler producing the report can
provide additional information, such as the offset from the allocation start,
the stack trace of deallocation (for heap-use-after-free), and the stack trace
of allocation (for both classes of bugs). Where possible, the report should
indicate if the erroneous access is a read or write.

This makes GWP-ASan error messages more informative than other typical crash
reports. In other words, GWP-ASan is capable of producing the same quality of
reports as ASan or HWASan. Listings~\ref{lst:uafreport} and \ref{lst:oobreport}
show examples of GWP-ASan reports.\footnote{The examples are modified to fit
this document's layout. All implementations in \S\ref{sec:impls} display
variations of these example reports.}

\begin{lstlisting}[frame=single, caption=Example GWP-ASan use-after-free report.,
  label=lst:uafreport, basicstyle=\footnotesize]
*** GWP-ASan detected a memory error ***
Use-after-free write at 0x7feccab26008 by thread 31027:
  #1 ./test(foo+0x45) [0x55585c0afa55]
  #2 ./test(main+0x9f) [0x55585c0af7cf]

The access is within 41B allocation at 0x7feccab26000

0x7feccab26000 was deallocated by thread 31027:
  #1 ./test(main+0x83) [0x55585c0af7b3]

0x7feccab26000 was allocated by thread 31027:
  #1 ./test(main+0x57) [0x55585c0af787]
*** End GWP-ASan report ***
\end{lstlisting}

\begin{lstlisting}[frame=single, caption=Example GWP-ASan out-of-bounds report.,
  label=lst:oobreport, basicstyle=\footnotesize]
*** GWP-ASan detected a memory error ***
Out-of-bounds read at 0x7feccab25ffe by thread 31027:
  #1 ./test(foo+0x45) [0x55585c0afa55]
  #2 ./test(main+0x9f) [0x55585c0af7cf]

The access is 2B left of 41B allocation at 0x7feccab26000

0x7feccab26000 was allocated by thread 31027:
  #1 ./test(main+0x57) [0x55585c0af787]
*** End GWP-ASan report ***
\end{lstlisting}

\section{Implementations}
\label{sec:impls}

This section lists various existing variants of GWP-ASan and some of their most
notable features.

\subsection{TCMalloc}

TCMalloc~\cite{TCMalloc} is the open-source \inlcode{malloc()} implementation
used in Google server-side code. It is highly optimized for CPU and RAM
efficiency on large multi-threaded applications.  It contains the first
(historically) implementation of GWP-ASan.

TCMalloc's implementation of \inlcode{WantToSample()} did not introduce a single new
instruction on the hot path of the allocator. This was made possible by
piggybacking on the existing sampling mechanism in TCMalloc used for heap
profiling. Effectively, the GWP-ASan sampling logic hides behind
pre-existing sampling logic.

\inlcode{IsGuarded()} is implemented as a bit test on the pointer value, thus
requiring zero memory loads.

\subsection{Google Chrome}
\label{sec:chrome}

Chrome implements a custom version of GWP-ASan, which hooks \inlcode{malloc()}
using Chrome's unified \inlcode{malloc()} shim that works on all of Chrome's
supported platforms.\footnote{Open source as part of the Chromium project:
\url{https://chromium.googlesource.com/chromium/src/+/lkgr/docs/gwp\_asan.md}}
The \inlcode{malloc()} shim requires an indirect call in every process using
GWP-ASan's hook.  This hook uses the simple \inlcode{WantToSample()}
implementation above.

For stability and security reasons, Chrome has a multiprocess architecture,
including a main browser process, a GPU process that renders to the screen, and
a group of many ``renderers'' that run websites or tabs. GWP-ASan is randomly
enabled in each process at process startup time, with a small probability in
frequently-launched renderer processes, and a larger probability in the
unsandboxed security-sensitive browser process, of which only a single instance
exists. If not enabled this avoids the runtime overhead of the indirect call
and allows GWP-ASan to use more memory per enabled process. It also prevents
some user frustration if a frequently occurring bug is causing many GWP-ASan
crashes.

Chrome's GWP-ASan uses the simple implementation described above, with the
exception that the maximum number of simultaneously allocated slots is set to
be significantly smaller than the total number of reserved allocation slots
(constant \inlcode{kReservedSlots}). This setup delays the reallocation of each slot,
forming a quarantine, while limiting the amount of physical memory overhead.
Finally, because each slot is associated with out-of-line memory-hogging
metadata such as compressed stack traces, the total number of slots associated
with metadata is selected to be much smaller than \inlcode{kReservedSlots}. If
\inlcode{kReservedSlots} were lower the quarantine would be much less effective.
The downside is that if a UAF occurs on a slot that has lost its metadata, the
resulting bug report is far less detailed and actionable.

Because GWP-ASan is an alternative allocator, it must provide at least the same
security guarantees as Chrome's hardened allocator, PartitionAlloc. This
includes never allocating object types from separate ``partitions'' in the same
slot for the lifetime of the process (to avoid simple type confusion exploits)
and never reusing a memory address as long as there are known references to the
object that once resided at that address~\cite{MiraclePtr}.

\subsection{Android/LLVM}

The \emph{compiler-rt} project~\cite{compiler-rt}, which exists within
LLVM~\cite{LattnerA2004}, contains an implementation of GWP-ASan that's used
for both Android and the Scudo hardened allocator.\footnote{Open source
documentation available at: \url{https://llvm.org/docs/GwpAsan.html}} This
implementation of GWP-ASan is designed to be used as a library, and can be
integrated with any allocator through a simple hook in \inlcode{malloc()} and
\inlcode{free()}.

Android, by default, has been using GWP-ASan for its system processes since
Android 11 (September 2020), and apps have historically been opt-in. As of
Android 14 though, apps will have GWP-ASan enabled by default (with an
opt-out), but in what we call a ``recoverable'' mode. In this mode, apps will get
a full GWP-ASan crash report, with stack traces and an error description, but
the app won't be forced to crash with a segmentation fault. Instead, the
system's signal handler (libsigchain, which runs before any signal handler
installed by the app) will also disable GWP-ASan, make the faulting page
read/write-able, and retry the faulting instruction. Use-after-free and
buffer-overflow will thus succeed in writing to a wrong memory location on the
restart, and reads will return zero. The primary benefit here is app
compatibility---the Android app ecosystem contains large quantities of apps
with memory-unsafe native code, and in the interest of user experience, such
apps should not be forced to crash when an Android user updates their device to
Android 14.

An additional Android-specific challenge is memory consumption. Unlike server
binaries, where hundreds of kilobytes is unnoticeable, Android has many
processes (\({\sim}200\)) running simultaneously, and runs on
memory-constrained devices. To reduce memory pressure, we apply two techniques:

\begin{enumerate}

  \item \runinsec{Metadata compression.} The largest part of GWP-ASan's
  metadata are the stack traces, the LLVM implementation of GWP-ASan encodes
  each stack frame as the ULEB-encoded difference between the return address of
  the current frame, and the frame before it. This provides a 50-75\% decrease
  in memory usage of storing stack traces.

  \item \runinsec{Process sampling.} We introduce another layer of sampling,
  and only turn on GWP-ASan for \(1/128\) process launches.

\end{enumerate}

These two techniques, combined with a carefully tuned small pool of allocations
(only 16 at any one time), means that an average device is only using
\({\sim}140\) KiB of extra memory for GWP-ASan (\({\sim}70\) KiB per sampled
process).  In addition, like in Chrome, the process sampling prevents frequent
crashes on the same process, which can lead to user frustration.

\subsection{Firefox}

Firefox implements its own custom version of GWP-ASan, named Probabilistic Heap
Checker (PHC).\footnote{PHC is open source, available at:
\url{https://searchfox.org/mozilla-central/source/memory/replace/phc/PHC.cpp}}
PHC is closely related to Chrome's GWP-ASan (\S\ref{sec:chrome}). The internal
Firefox allocator \emph{mozjemalloc} offers \inlcode{malloc()}-replace support,
allowing the PHC implementation to intercept the respective allocator
functionality and handle PHC-controlled allocations separately.

Since Firefox has a different allocation profile compared to Chrome (both in
frequency and allocation size), PHC uses slightly different parameters to
decide when and how to sample allocations. This is particularly relevant to
decide when to start sampling at process startup.

\subsection{Apple Platforms}

Apple's variant of GWP-ASan, named Probabilistic Guard Malloc~(PGM), is
implemented in the standard user space allocator.\footnote{PGM is part of Apple
libmalloc:
\url{https://github.com/apple-oss-distributions/libmalloc/blob/main/src/pgm\_malloc.c}}
It was first deployed to customer populations with iOS~14.5 and macOS~11.3
(April~2021) and deployment gradually expanded to additional platforms,
including watchOS and tvOS.  PGM is enabled for all Apple-owned user space
processes (including apps) and integrates with the existing crash reporting
pipeline.  Crash reports are augmented with additional information about the
guarded allocation, most notably the allocation and deallocation stack traces.
PGM does not apply to third-party apps and there are a small number of
processes that use a custom memory allocator for some or all of their
allocations.  Since these allocations are served by a different allocator
implementation, they do not benefit from PGM.

PGM uses conservative sampling rates and a fixed per-process memory budget to
ensure performance remains unaffected.  The per-process memory budget is 2~MiB
(except for macOS, where it is 8~MiB) and bounds PGM's total memory footprint.
It accounts for all reserved VM pages (guard, allocation, and quarantine pages)
and allocation metadata.  Stack traces are stored in compressed form.
\emph{Process sampling} is used to ensure the number of simultaneously
protected processes on a device is very small.  The average number of protected
processes per device is tuned to be~0.5.  In addition to tuning system-wide
overhead, process sampling also limits user impact by avoiding crash loops.

\subsection{Linux Kernel}

The Linux kernel (since version 5.12) has its own implementation, named Kernel
Electric-Fence (KFENCE).\footnote{KFENCE is part of the mainline Linux kernel:
\url{https://docs.kernel.org/dev-tools/kfence.html}} KFENCE's implementation
has to integrate with the Linux kernel SLAB and SLUB heap
allocators~\cite{SLUB}, with the latter having become the recommended default
allocator of the Linux kernel.

By virtue of being implemented in an OS kernel, KFENCE's implementation is more
complex. Several abstractions that user space can rely on are unavailable in an
OS kernel environment (signals, page-protection). Instead, per-target
architecture support is required to deal with both page-fault handling and page
protection. Indeed, page-protection from user space (using a default
\inlcode{mprotect()}) is rather costly, especially due to TLB shootdowns.
Instead, on some architectures (such as X86), KFENCE implements lazy
page-protection, where only the local TLB is invalidated, with the only
downside being few missed bugs (viz. false negatives).

Another significant difference in KFENCE's implementation is that its decision
to sample allocations is time-based: the implementation uses a fixed sample
interval (with a default of 100 milliseconds) and a timer to toggle a bit
checked by the SLAB or SLUB allocators. The kernel's allocator pressure is
entirely workload dependent and hard to predict. Since the kernel has to serve
any number of different workloads at different points in time or even
concurrently for the entire uptime of the system, KFENCE's decision to sample
should be independent of potentially pathological or even malicious workloads,
yet remain predictable to derive reliable performance characteristics of the
system with KFENCE enabled. A time-based sampling policy is independent of
allocator pressure, relatively uniform, and therefore provides for a
predictable upper bound on kernel-wide sample rate.

The current default implementation checks a simple boolean ``gate'' in the
allocator fast path, which results in a load, compare, and conditional jump:
this implementation has been measured to have negligible impact across a
variety of real-world production workloads. We note that an initial version of
KFENCE used a dynamically patched branch (self-modifying code) to avoid
checking a boolean in the fast path at all. In certain configurations with
relatively large sample intervals of more than 500 milliseconds, this can
minimally perform better, especially on systems with few CPUs (dual or quad
core systems). On larger systems with large CPU counts, however, patching the
branch using the kernel's existing code-patching machinery (called "Static
Keys"~\cite{LinuxStaticKeys}) is equivalent to taking a global system lock
which turned out to be unacceptable.

\subsubsection{\textbf{Optimizing KFENCE Pool Utilization}}
\label{sec:kfence_bloom}

KFENCE uses a fixed pool of object pages and adjacent guard pages that are set
up once on boot. This pool must be able to service KFENCE allocation requests
until the next reboot of a system. One problem with that is dealing with
long-lived allocations eventually consuming the entire pool: we implement a
policy that rejects new allocations if they are unlikely to contribute to new
\emph{coverage}. We define the \emph{coverage source} of an allocation to be
based on its allocation stack trace. More specifically, if pool utilization
reaches 75\% or above, KFENCE skips new allocations if an existing valid
allocation of the same coverage source exists. The implementation hashes the
stack trace and a Counting Bloom filter is used to efficiently query if one or
more allocations of the same coverage source exist. The policy ensures diverse
coverage of allocations, and as a side-effect limits frequent long-lived
allocations (e.g.  filesystem caches).

%% file: tex/eval.tex
\newcommand{\figCrashes}{%
  \begin{figure*}[t!]
    \centering
    \begin{tikzpicture}
      \node (img1)  {\includegraphics[width=0.95\linewidth]{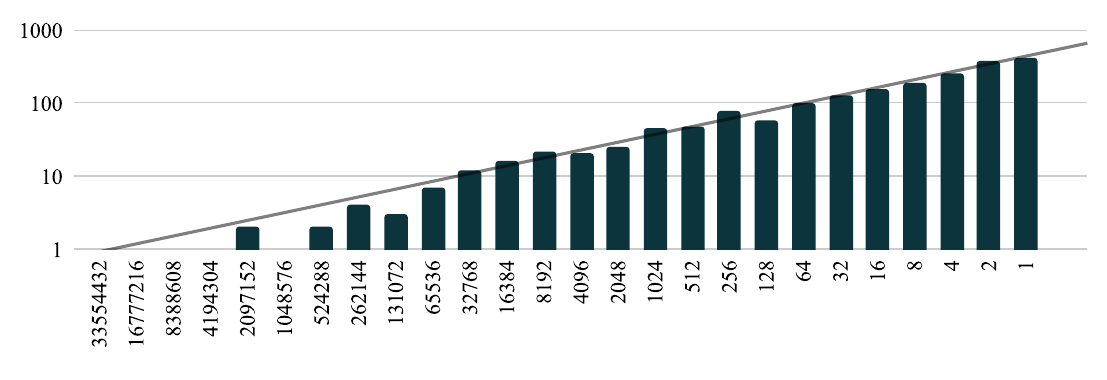}};
      \node[below=of img1, node distance=0cm, yshift=1.7cm] {Number of GWP-ASan crash reports received ($2^n$)};
      \node[left=of img1, node distance=0cm, rotate=90,anchor=center,yshift=-1.3cm] {Number of unique bugs ($10^n$)};
    \end{tikzpicture}
    \vspace{-1.2em}
    \caption{Bug occurrences across Google server-side applications, Android, and
    Chrome.}
    \label{fig:crashes}
  \end{figure*}
}

\section{Results}
\label{sec:results}

This section discusses the results of real-world deployment of GWP-ASan
variants and our experience over the past several years.

\subsection{Google Server-Side Software}

GWP-ASan for the Google server-side software was the first variant we deployed,
with the first production bugs observed in late 2018. In 2019, 300+ bugs were
reported and fixed. In 2020 and 2021 we observed 450+ fixed bugs annually, and
500+ bugs in 2022. The trend continued in 2023 with a total of 550+ bugs fixed.

Approximately 80\% of the fixed bugs were heap-use-after frees, and 20\% were
heap-buffer-overflows.

Since 2019 100+ bugs were marked as ``can't reproduce'', which would typically
indicate that the developers did not have sufficient information to understand
and fix the bug. 400+ bugs were marked as ``obsolete'', which would typically
indicate that the bug is unimportant for some reason (e.g. the code is deleted
or represents a one-time experiment).

The GWP-ASan reports are processed by the telemetry and bug reporting systems
in the same way as any other process crashes. The only notable difference is
that GWP-ASan reports have two (for buffer overflows) or three (for use after
free) stack traces, while the majority of other crashes (e.g. NULL
dereferences) have only one stack trace.

We have evaluated GWP-ASan performance using performance-sensitive load tests
and production telemetry~\cite{RenTMSRH2010}.  With the selected sampling rate,
the performance impact is effectively undetectable.

There were several cases where GWP-ASan detected and reported that the root
cause of an ongoing issue in a production service was caused by a
use-after-free or heap-buffer-overflow.

\subsection{Google Chrome}

GWP-ASan has been enabled by default on Windows and Mac since Chrome 80 in
2019, and has been partially enabled on Linux and ChromeOS since 2022.

Bug reports are prescreened by security team members before filing them in the
bug tracker, as GWP-ASan sometimes produces inactionable reports. A faulty
access to GWP-ASan's guarded region may occur long after another memory-safety
error has occurred and the resulting bug report appears nonsensical. Some
methods can be used to reduce or eliminate some classes of common inactionable
reports (e.g. those that occur when ASCII values happen to form a valid
pointer to a guarded allocation.) In the last 120 days (as of Jun 07 2023),
71\% of crashes are reported as bugs after the prescreening (after filtering
for crashes in old system-provided libraries and third-party code).

There have been 271 bugs
filed,\footnote{\url{https://crbug.com?q=Hotlist\%3DGWP-ASan\&can=1}}
of which 176
(65\%)\footnote{\url{https://crbug.com?q=Hotlist\%3DGWP-ASan\%20Type\%3DBug-Security\&can=1}}
have been judged to be possibly exploitable by attackers.

Of the 243 bugs that have been marked as
resolved,\footnote{\url{https://crbug.com?q=Hotlist\%3DGWP-ASan\%20status\%3AFixed\%2CVerified\%2CDuplicate\%2CWontFix\%2CArchived\&can=1}}
168 (69\%) have been marked as
fixed.\footnote{\url{https://crbug.com?q=Hotlist\%3DGWP-ASan\%20status\%3AFixed\%2CVerified\&can=1}}
35 (14.4\%) have been marked as
``WontFix'',\footnote{\url{https://crbug.com?q=Hotlist\%3DGWP-ASan\%20status\%3AWontFix\&can=1}}
meaning the developer has judged the bug to not be actionable in any way. Other
resolutions include marking the bug as a duplicate, or marking it as the
responsibility of an external dependency, for example, bugs in macOS system
code, for which we receive crash reports when they occur in a Chrome process.

180 of the bugs are
heap-use-after-free,\footnote{\url{https://crbug.com?q=Hotlist\%3DGWP-ASan\%20heap-use-after-free\&can=1}}
and 35 are
heap-buffer-overflow.\footnote{\url{https://crbug.com?q=Hotlist\%3DGWP-ASan\%20heap-buffer-overflow\&can=1}}
21/32 (65.6\%) of the resolved heap-buffer overflows are marked as fixed or a
duplicate of an existing
issue,\footnote{\url{https://crbug.com?q=Hotlist\%3DGWP-ASan\%20heap-buffer-overflow\&can=1}}$^,$\footnote{\url{https://crbug.com?q=Hotlist\%3DGWP-ASan\%20heap-buffer-overflow\%20status\%3AFixed\%2CVerified\%2CDuplicate\&can=1}}
and 148/167 (88.6\%) of the resolved heap use-after-frees are marked as fixed
or
duplicate.\footnote{\url{https://crbug.com?q=Hotlist\%3DGWP-ASan\%20heap-use-after-free\%20status\%3AFixed\%2CVerified\%2CDuplicate\%2CWontFix\%2CArchived\&can=1}}$^,$\footnote{\url{https://crbug.com?q=Hotlist\%3DGWP-ASan\%20heap-use-after-free\%20status\%3AFixed\%2CVerified\%2CDuplicate\&can=1}}

218/271
(80.4\%)\footnote{\url{https://crbug.com?q=Hotlist\%3DGWP-ASan\%20\%22reported\%20by\%20GWP-ASan\%22\&can=1}}
of the GWP-ASan bugs filed were found and reported by GWP-ASan before any other
crash bug was filed for the same crash. This is likely because GWP-ASan bugs
can be filed when only one crash has occurred (subject to prescreening),
whereas regular crash bugs are typically inactionable with only a single
report.

\subsection{Android}

GWP-ASan has been enabled by default for system processes and system apps since
Android 11. Since then, we've detected and fixed a large quantity of both
previously undetected memory-safety bugs, and new regressions.

Within 60 days (starting in May 2023), we detected 1,972 unique stack traces in
system processes across the Android ecosystem from ${\sim}11.7M$ crash reports.
We consider a stack trace to be unique when the function name (where available)
or dynamic shared object (DSO) from the most significant frame of the bad
memory access (excluding common frames like libc or Android Runtime), is
unique.

We believe that a small fraction of these crashes are caused by hardware
failures and not by software bugs, however, reports observed more than once
clearly indicate real memory-safety bugs: 54\% (1,066) of unique stack traces
were observed 5 or more times.
Of these crashes, 57\% were use-after-free, 27\% were buffer-overflow, 4\% were
buffer-underflow, and the remainder were double free, invalid free, or
indeterminate bugs.

\figCrashes

\subsection{Firefox}

To understand PHC results on Firefox, there are two important aspects to
consider. First, PHC is currently only running on the \emph{Nightly}
channel and therefore has limited user exposure. We are working on bringing PHC
to Firefox Release by the end of 2023. Second, prior to PHC the ``ASan Nightly
Project'' was used on Firefox. In this project, a Firefox Nightly version built
with AddressSanitizer was shipped to users who manually opted in to this (much
slower) version. Although the user base was of course limited, ASan Nightly did
not rely on probabilistic sampling and therefore it found a significant chunk
of existing use-after-free issues. Overall we assume that this project is
responsible for eliminating a lot of the long-standing bugs that we would have
found otherwise at the beginning of PHC deployment. It is important to note
that results from the ASan Nightly Project directly influenced the decision to
implement PHC: the project showed us that use-after-free issues can be fixed
almost 100\% of the time if the developers have the ``free'' stack trace in
addition to the usual ``use'' stack trace.

To date, PHC in Firefox has uncovered 7 use-after-free issues and 2
buffer-overflows, some of which were in third-party code and fixed upstream.
For some of these bugs, we also saw collisions with reports from other sources
(e.g. the ASan Nightly Project). We expect that deployment to more clients
through the Release channel will further increase the detection rate and allow
us to identify regressions more quickly.

\subsection{Apple Platforms}

The Apple crash reporting pipeline groups crashes using the notion of an
\emph{issue signature}.  Conceptually, an issue signature consists of the stack
trace of the crashing thread.  For Probabilistic Guard Malloc~(PGM) crashes,
the system files a bug for every new pair of process and issue signature (see
\S\ref{sec:occurrences_per_bug}).

As of September~2023, a total of 3,748 PGM bugs have been filed of which 1,438
are marked fixed with an associated code change.  Of all bug reports, only 13
were closed without a resolution, meaning that despite the additional
information in the crash report component owners were unable to diagnose the
bug without a reproducer.  This compares very favorably with standard crash
reports for memory errors, resulting in a 99\% fix rate for PGM bugs.  Another
27 bugs were found in local experiments and code that was recently removed.
The remaining 2270 bugs were marked as duplicates resulting from PGM finding
already-fixed bugs and finding the same bug in code shared by multiple
processes (bugs are filed for pairs of process and issue signature).

Of the found bugs, 77\% constitute use-after-free bugs and 23\% are
buffer-overflows.  For about a third of fixed bugs the diagnosed root cause was
a concurrency or thread safety issue.  Interestingly, there even was an example
for a locally-reproducible bug that was detected using PGM, but eluded
detection by AddressSanitizer. Our explanation is that PGM's lower overhead
was more conducive to reproducing the error condition.

In summary, PGM has been an effective tool for finding and diagnosing memory
errors at Apple.  On average, 2.1 new bugs have been found \emph{every day}
since it was first deployed at scale in April~2021.  The additional information
in PGM crash reports (most notably, allocation and deallocation stack traces)
makes them actionable even without a reproducer, resulting in a high 99\% fix
rate.  In a handful of cases, a single PGM crash report made the difference for
diagnosing a known high-impact bug.  PGM even found bugs (now fixed) in code
that had remained unchanged for over 20 years.

\subsection{Meta}

Meta has used a variant of GWP-ASan in the Facebook and Messenger Apps on
Android since 2020. GWP-ASan is bundled with the applications in a way that
allows for it to be dynamically configured and tuned for a device population.
This has allowed Meta to ship and enable GWP-ASan on a wide range of Android
devices and versions, control key allocator properties such as the sampling
rate, and an allow-list of modules subject to allocation sampling. The default
sampling rate for a process is 1/1000 with a memory budget of 1024 object
pages (8 MiB). Meta can additionally opt devices in or out of sampling at
application launch time based on eligibility criteria such as available system
memory.

When a sampled allocation is determined to have resulted in a memory-safety
issue, a crash report is generated containing custom streams with relevant
information from the sampling such as allocation and deallocation stack traces.
Reports are ingested by Meta infrastructure as part of the regular crash
reporting pipeline and integrated directly with crash analysis tools that
engineers already frequently use. Integration of GWP-ASan in their stack has
over time helped Meta fix multiple difficult-to-debug reliability issues caused
by use-after-frees and buffer overflows.

\subsection{Linux Kernel}

KFENCE has been deployed in Google's server kernels, Android, and ChromeOS
kernels. We're also aware of numerous third parties using KFENCE, given it
works out of the box by simply changing a kernel build-time configuration
parameter. More effort is required to collect and analyze KFENCE reports from a
fleet of machines, which needs additional tooling. To date, KFENCE has reported
60+ bugs in Google's downstream Linux kernels. The upstream Linux kernel up to
version 6.3 has 12 fix commits mentioning KFENCE-reported bugs since the
introduction of KFENCE in 2021 with Linux 5.12.

\subsection{Analysis of Occurrences Per Bug}
\label{sec:occurrences_per_bug}

The number of occurrences of every bug detected by GWP-ASan in production
deserves a separate discussion. Intuitively, bugs that occur in production
frequently will also be frequently reported with a sampling-based bug detector.
But we don't know the frequency of the bug occurrences, only their detection
frequency.

We analyzed the detection frequency in the Google server-side applications, the
Android platform (both the user space and the kernel), and Chrome, shown in
Figure~\ref{fig:crashes}. In all cases, we observed a very similar picture.
Roughly half of the bugs are only ever seen once, a quarter of the bugs are
seen 2-10 times, and very few bugs are detected hundreds of times. The log-log
plot looks like a straight line, which is frequent for such phenomena.

Because of the low sampling rate GWP-ASan uses, most bugs are only detected
once. We suspect that many more bugs occur in production infrequently enough to
have never been detected by GWP-ASan.

%% file: tex/related.tex
\section{Related Work}
\label{sec:related}

As discussed in the introduction, the idea to rely on page protection and
dedicated guard pages around the object page for memory-safety error detection
was first found in the Electric Fence Malloc Debugger~\cite{EFence} for various
POSIX-compatible operating systems. For Microsoft Windows,
PageHeap~\cite{PageHeap} serves a similar purpose and can find memory-safety
errors. To the best of our knowledge, PageHeap has a sampled mode but was never
intended for production use due to large overheads. The Debug Malloc
Library~\cite{DMalloc} is another allocator replacement, giving developers a
library to track object state to help debug errors; however, its overheads also
make it unsuitable for production.

While sampling program executions to analyze system performance at scale has
become ubiquitous~\cite{RenTMSRH2010}, sampling to detect errors is less
common. Liblit et al.~\cite{LiblitAZJ2003} first proposed compile-time
assertion sampling to detect bugs, including memory safety bugs. A large number
of builds for a single application are made, where different builds have a
small subset of assertions enabled; instrumentation with
CCured~\cite{NeculaMW2002} is used as an ``assertion source'' to find memory
safety bugs. A benefit of this approach is that it is not tied to any
particular compile-time instrumentation, and could be combined with newer
state-of-the art compiler-instrumentation based dynamic analysis, such as
HWASan~\cite{SerebryanySSTV2018}. Cooperative Bug Isolation
(CBI)~\cite{NainarCRL2007} applies compiler instrumentation to collect values
of simple predicates observed at run-time; a post-execution statistical
analysis is performed to find anomalies and predict bugs. Here as well,
compile-time sampling is used to reduce the instrumentation overhead.  These
approaches have at least two challenges: (a) maintaining multiple builds for a
single application may be expensive at scale and (b) the overhead of
compiler-instrumented code is still non-zero and if a hot function happens to
be instrumented in a given build, the slowdown may be significant.

Hauswirth et al.~\cite{HauswirthC2004} propose using a binary rewriting system
to instrument the code for bug detection and a dynamic dispatch to choose
between instrumented and non-instrumented code in order to sample the checks at
run-time. The drawback is that the dynamic dispatch code is not free. The paper
describes the idea of ``sampling executions of code segments at a rate
inversely proportional to their execution frequency,'' which remains promising.

%% file: tex/future.tex
\section{Future Work}
\label{sec:future}

Assuming memory unsafe code will remain in wide use and pre-production testing
will remain imperfect, we need to keep improving tools like GWP-ASan. Some
notable directions are:

\begin{enumerate}

  \item Extend GWP-ASan or explore new low-overhead sampling-based bug
    detection algorithms for bug classes currently not found by GWP-ASan. Of
    particular interest are stack-use-after-return bugs: our estimate based on
    pre-production testing is that these bugs are roughly 40\% as frequent as
    heap-use-after-free bugs; their frequency in production is unknown to us.
    Furthermore, too many concurrency issues escape to production, and we
    expect solutions for low-overhead sampling-based data-race detection to be
    profitable~\cite{GWPTsan}.

  \item Tune the existing implementations to skew towards less frequent heap
    allocation sites, similar to what KFENCE does (\S\ref{sec:kfence_bloom}).
    This assumes that frequent allocations are better tested anyway.

  \item Find mechanisms that allow higher sampling rates. This may include
    improving the scalability of the mmap system call in operating system
    kernels, or using hardware features such as Intel Memory Protection Keys
    (MPK)~\cite{PKU} or Arm Memory Tagging (MTE)~\cite{Serebryany2019}.

  \item Combine with other related detection mechanisms, e.g. Chrome's
    lightweight use-after-free detector which has a higher sampling rate but
    does not detect all types of use-after-frees~\cite{LightweightUafDetector}.

  \item Ensure that major implementations can handle allocations of any size.

  \item Create a feedback loop from earlier executions to newer ones. For
    example, by increasing sampling rates for allocations previously involved
    in a bug report, and decreasing sampling rates for long-lived allocations
    that are less likely to cause bugs.

  \item Create mechanisms to dynamically direct the sampling budget towards a
    specific process to help project teams track down hard-to-diagnose bugs
    that evaded all other forms of testing.

\end{enumerate}

%% file: tex/conclusion.tex
\section{Conclusion}

Memory safety remains a major unresolved problem. The industry must migrate
away from memory-unsafe code, but this will take decades. In the meantime,
tools like GWP-ASan offer a low-overhead and easy-to-deploy option for bug
detection in production. These tools rely on telemetry systems that send crash
reports to developers. They must be used along with available pre-production
detection mechanisms and security mitigations.

GWP-ASan is not a security mitigation mechanism. When used, however, it
improves the overall product security by allowing developers to detect and fix
many vulnerabilities.

\section*{Acknowledgements}

We would like to thank our colleagues and the respective open-source
communities for their helpful feedback, reviews, and comments.  We would like
to thank the anonymous reviewers for their helpful comments and advice.